# Spatial Organization in the Reaction A + B → inert for Particles with a Drift

S. A. Janowsky
Department of Mathematics
University of Texas
Austin, TX 78712
janowsky@math.utexas.edu

**Abstract**

We describe the spatial structure of particles in the (one dimensional) two-species annihilation reaction A + B → 0, where both species have a uniform drift in the same direction and like species have a hard core exclusion. For the case of equal initial concentration, at long times, there are three relevant length scales: the typical distance between similar (neighboring) particles, the typical distance between dissimilar (neighboring) particles, and the typical size of a cluster of one type of particles. These length scales are found to be generically different than that found for particles without a drift.

## 1 Introduction

In the irreversible two species annihilation reaction A + B → 0, local fluctuations in the particle density are sufficiently important that (at least in low dimension) the mean field rate equations do not accurately predict the decay of the density when the initial concentrations of the two species are equal [1, 2, 3, 4, 5, 6, 7, 8]

There is a standard picture for what does happen [1]: in a region of size $L$ it typically takes a time on the order of $L^2$ for all of the different particles to react, since they may need to diffuse across the entire region in order to annihilate. One species in this region will be completely eliminated and excess particles of the other species will remain; their number is the initial excess of either type A or type B particles, and its expected value is proportional to the square root of the volume. At time $t$ the concentration $c(t)$ thus behaves as

$$c(t) \sim \left(L^d\right)^{1/2} \Big/ L^d \sim \left(t^{d/2}\right)^{1/2} \Big/ t^{d/2} = t^{-d/4}, \tag{1}$$

for dimension $d \leq 4$, *i.e.* $t^{-1/4}$ in dimension one, which has been verified rigorously [4, 7].

The above result depends crucially on the fact that the particles undergo ordinary diffusion. Yet this might not be correct if there is a drift field, *even if both species drift in the same direction.* Clearly one expects to see different behavior if the two species drift in *opposite* directions: in addition to the diffusive length scale $L_\perp \sim t^{1/2}$ there will be a drift length scale $L_{||} \sim vt$; the simple heuristic reasoning that lead to eq. (1) in the drift-free case then yields [9, 10]

$$c(t) \sim \left(vt^{(d+1)/2}\right)^{1/2} \Big/ \left(vt^{(d+1)/2}\right) = v^{1/2} t^{-(d+1)/4} \tag{2}$$

for dimension $d \leq 3$, or $(v/t)^{1/2}$ in one dimension.

In a previous work [8], we claimed the more suprising result that even if the two species have a drift in the *same* direction, even if their relative velocity is zero, the result may still be different from the expected $t^{-d/4}$. This is because after subtracting off the average motion, there may be a super-diffusive length scale, of order $t^{2/3}$ in one dimension [11, 12]. Thus one might expect the concentration to go as

$$c(t) \sim L^{1/2}/L \sim \left[t^{2/3}\right]^{1/2} \Big/ t^{2/3} = t^{-1/3} \,. \tag{3}$$

This occurs when the long time behavior of the system is governed not by a linear diffusion equation, but by a *nonlinear* equation, the noisy Burgers equation [13],

$$\frac{\partial \rho}{\partial t} + \rho \frac{\partial \rho}{\partial x} = \nu \frac{\partial^2 \rho}{\partial x^2} + \frac{\partial \xi}{\partial x} \,, \tag{4}$$

where $\rho$ is a rescaled density, $\nu$ is the viscosity (representing the lattice spacing) and $\xi$ is a random noise term, *e.g.* white noise where the covariance is

$$\langle \xi(x,t) \xi(x',t') \rangle = 2\nu \delta(x - x') \delta(t - t') \,. \tag{5}$$

As an example of a system that falls into this universality class we choose the Asymmetric Simple Exclusion Process (ASEP). We thus consider 2 species of particles that individually undergo ASEP dynamics but annihilate when jumping on top of each other.

Both the drift *and* the exclusion rule are necessary for Burgers equation (4) to be relevant; a drift without the exclusion rule has no effect on the asymptotic behavior of the annihilation dynamics [14]. To keep matters simple we consider infinite drift—particles move either to the right or not at all. We use periodic boundary conditions on the one-dimensional lattice. A short description of the dynamics follows: at each time step, a site is picked at random. If that site is occupied, its particle attempts a jump to its nearest neighbor on the right. If the target site is empty, the jump succeeds. If the target site is occupied by a particle of the same species, the jump fails. If the target site is occupied by a particle of the opposite species, both particles annihilate.

The asymptotic behavior of the density in this system was studied in [8]. Here we provide further evidence that this system is in a different universality class than the system without a drift by examining the spatial organization of the reactants in this system. This type of analysis was done by Leyvraz and Redner [5, 6] for the drift-free model; our results are unambiguously distinct from theirs.

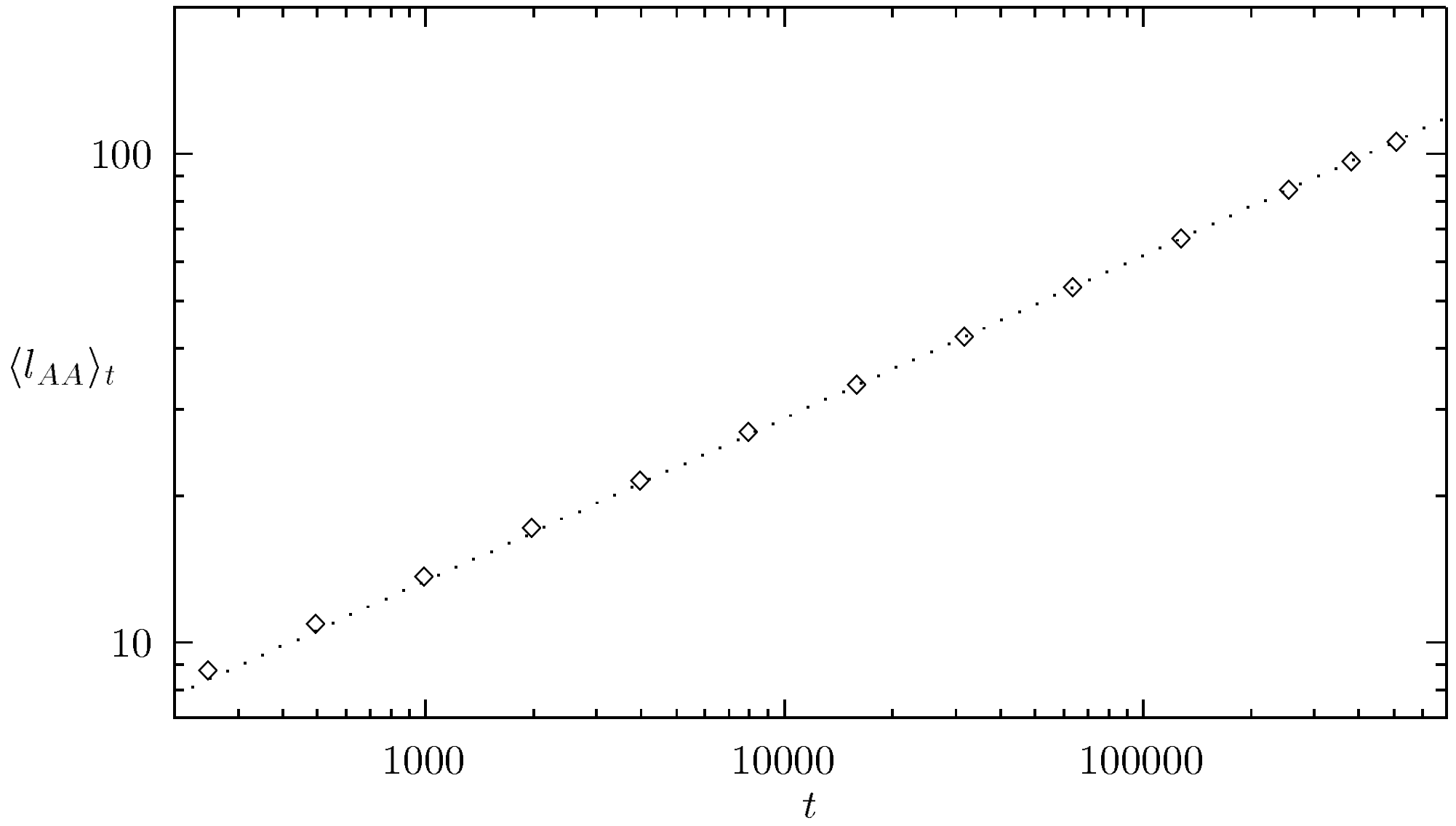

Figure 1: Log-log plot of the average same-species interparticle spacing, $\langle l_{AA} \rangle_t$. The dotted line has a slope of 1/3.

## 2 Distribution of interparticle distances

As in [5, 6], our most important results deal with three distributions: 1) $P_t(l_{AA} = x)$, the probability that nearest neighbor particles of the same type are at distance $x$ at time $t$; 2) $P_t(l_{AB} = x)$, the probability that nearest neighbor particles of different type are at distance $x$; 3) $P_t(L = x)$, the probability that a domain of one type of particle has length $x$. The length of a cluster $L$ is defined as the distance between the first particles of opposite species on either side of the cluster. We performed numerical simulations to measure these quantities; our results are based on averaging 9 independent runs of a system of size 4,000,000; times ranged up to 512,000 time steps. The initial density of each species was 0.45. (In [8] this density seemed to give the fastest approach to the asymptotic regime.)

We first consider $P_t(l_{AA} = x)$. For all times $t$, this appears to be a monotonically decreasing distribution, almost exponential in shape. The average same-species interparticle spacing increases with time, as is shown in figure 1; the average spacing increases as $t^{1/3}$. Recall that in the symmetric model the spacing grew as $t^{1/4}$. Here the interparticle spacing grows more rapidly; consistent with the more rapid decrease in density ($t^{-1/3}$) observed in [8]. It is interesting that the density took some time to reach the asymptotic regime, while the spacing seems to obey scaling even for very short times.

An example of the distribution itself is plotted in figure 2; the time here is 8000 but the graphs for other values of $t$ are similar. It appears to be the superposition of two exponential distributions— one with twice the decay rate of the other. The explanation for this is quite simple: when clusters coalesce, the new cluster that is formed is the sum of the lengths of the constituent clusters. Thus, neglecting correlations between neighboring clusters, the distribution is (approximately) proportional to

$$\exp(-x/\langle l_{AA}\rangle_t) + \text{ reaction rate } \times \exp(-x/\langle 2l_{AA}\rangle_t)\,, \tag{6}$$

since the clusters which come together can be of arbitrary size. This same type of distribution is seen in the symmetric case, except of course for the different behavior of the average spacing.

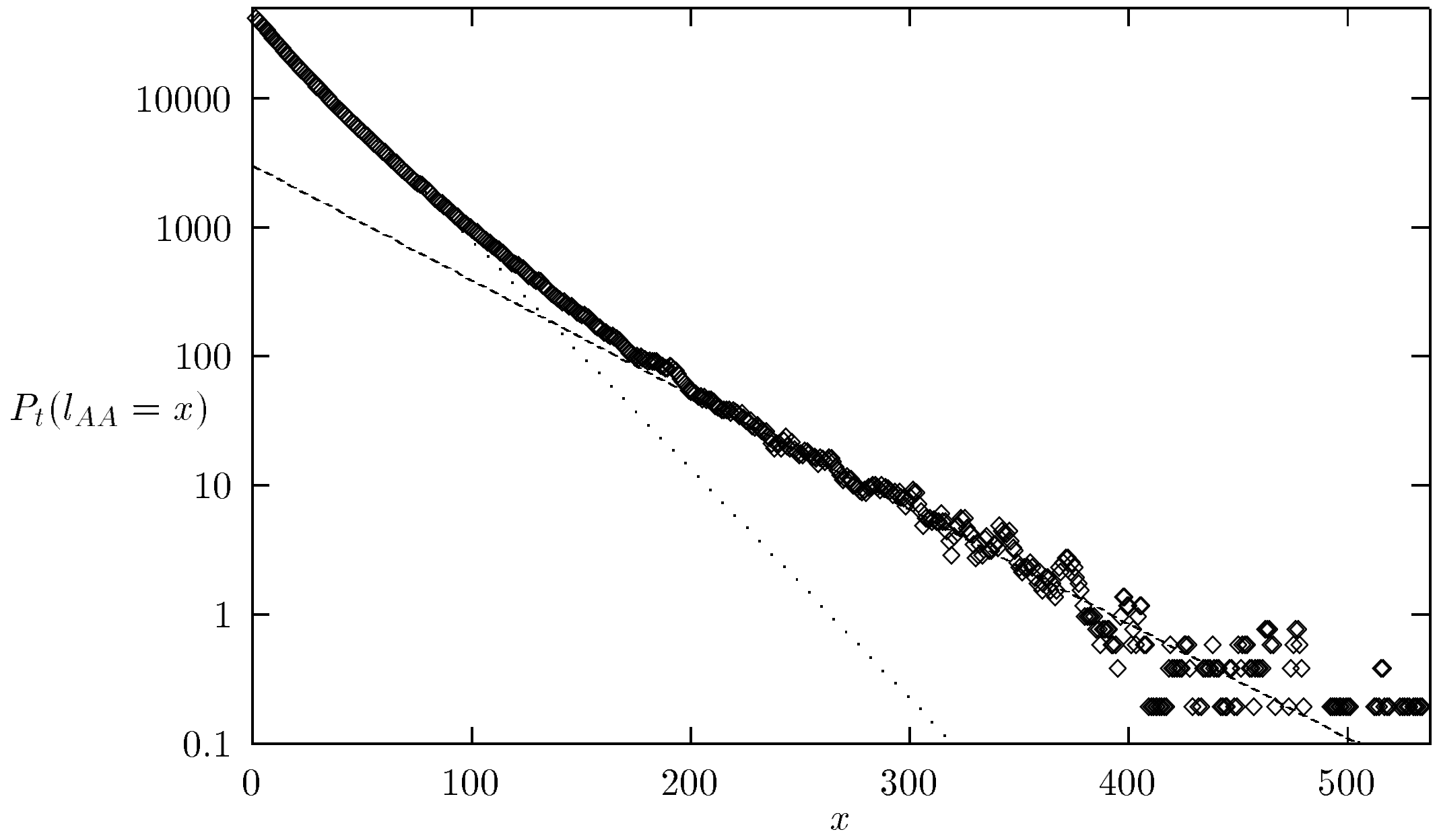

Figure 2: Semi-log plot of the (unnormalized) probability distribution for the same-species interparticle spacing at time 8000, $P_t(l_{AA}=x)$. The data is smoothed by averaging the distribution over 5 consecutive points. The slope of the steeper straight line is exactly double that of the other.

The distribution for the different-species interparticle spacing, or equivalently, the distance between domains, has a different shape entirely. Since A and B type particles annihilate if they come too close, there is an effective repulsion between particles of different species. Thus $P_t(l_{AB}=x)$ is zero at $x=0$, increases to a peak near $\langle l_{AA}\rangle_t$ (apparently linearly), and then decays (apparently exponentially); however, there does not seem to be a nice fit to the simple form $x\exp(-x/X_{AB})$ as is seen in the symmetric case [6].

Because of the effective repulsion, one expects that

$$\langle l_{AA}\rangle_t < \langle l_{AB}\rangle_t\,. \tag{7}$$

In fact we find that not only is the distance between unlike particles greater than that between like particles, but the growth rate is greater as well. Surprisingly, it appears to be the same growth rate as in the drift-free case: namely $\langle l_{AB} \rangle_t \sim t^{3/8}$, as is seen in figure 3.

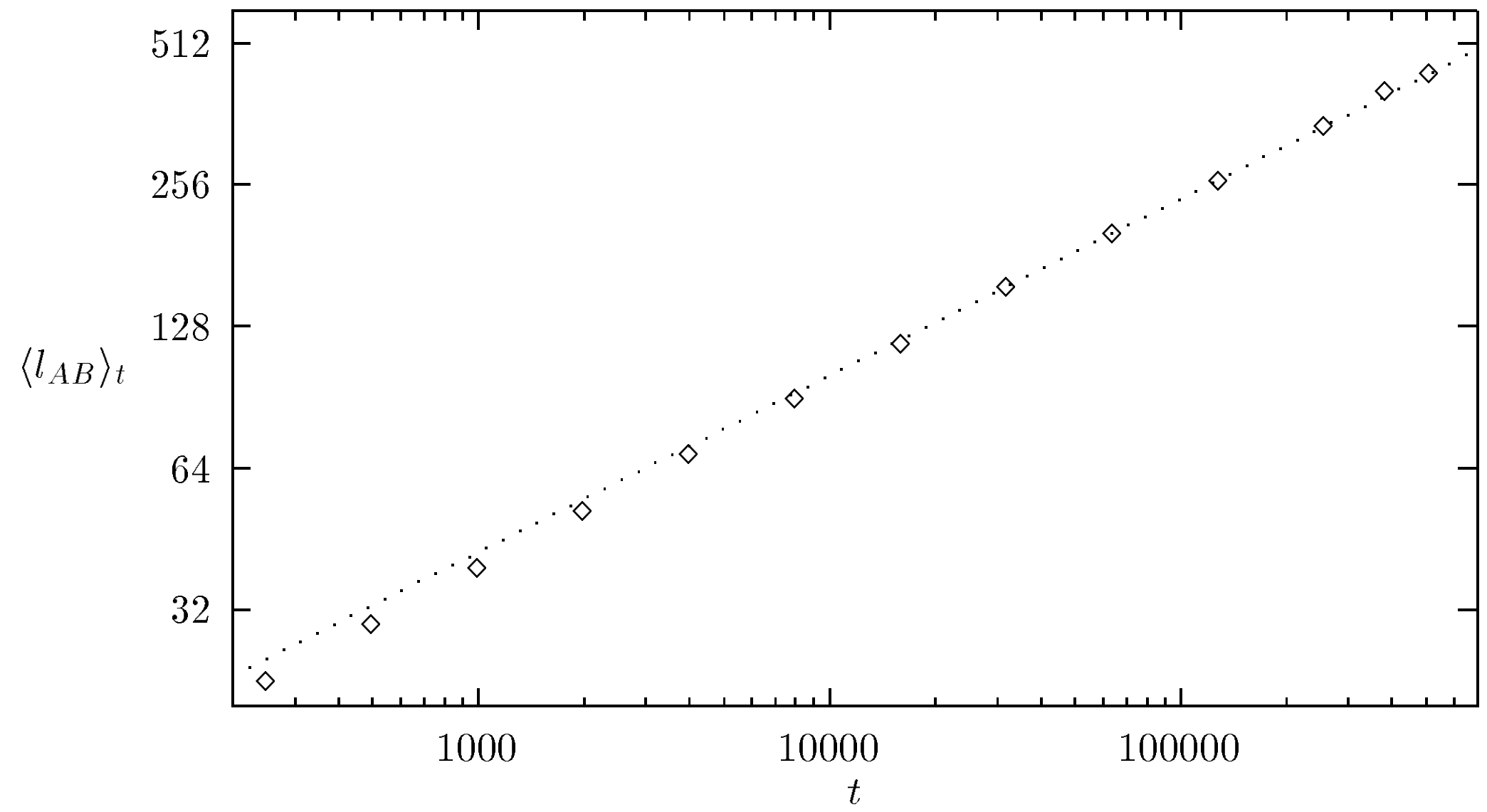

Figure 3: Log-log plot of the average different-species interparticle spacing, $\langle l_{AB} \rangle_t$. The dotted line has a slope of 3/8.

While 3/8, the opposite species growth rate, is only slightly larger than 1/3, the same species rate, this difference in length scales does mean that our clusters are in fact really clusters: inter-cluster distances are much larger than intra-cluster distances (at least for $t$ large); this validates the assumptions of [8] which are based on treating clusters as isolated from each other. The fact that the ratio of these lengths scales only as $t^{1/24}$ is perhaps responsible for the long times needed to observe asymptotic behavior for the density.

## 3 Particle distribution within a domain.

The third relevant length scale is the length of the clusters themselves. Since small clusters are quickly annihilated, and large clusters take a long time to form, we expect a single-peaked distribution for $P_t(L = x)$ that is zero at $x = 0$. In fact qualitatively the distribution looks very similar to $P_t(l_{AB} = x)$.

The average value for $L$ as a function of time is plotted in figure 4. As in the drift-free case, the slope of this line can be determined from the other length scales in the problem [6]. Basically the

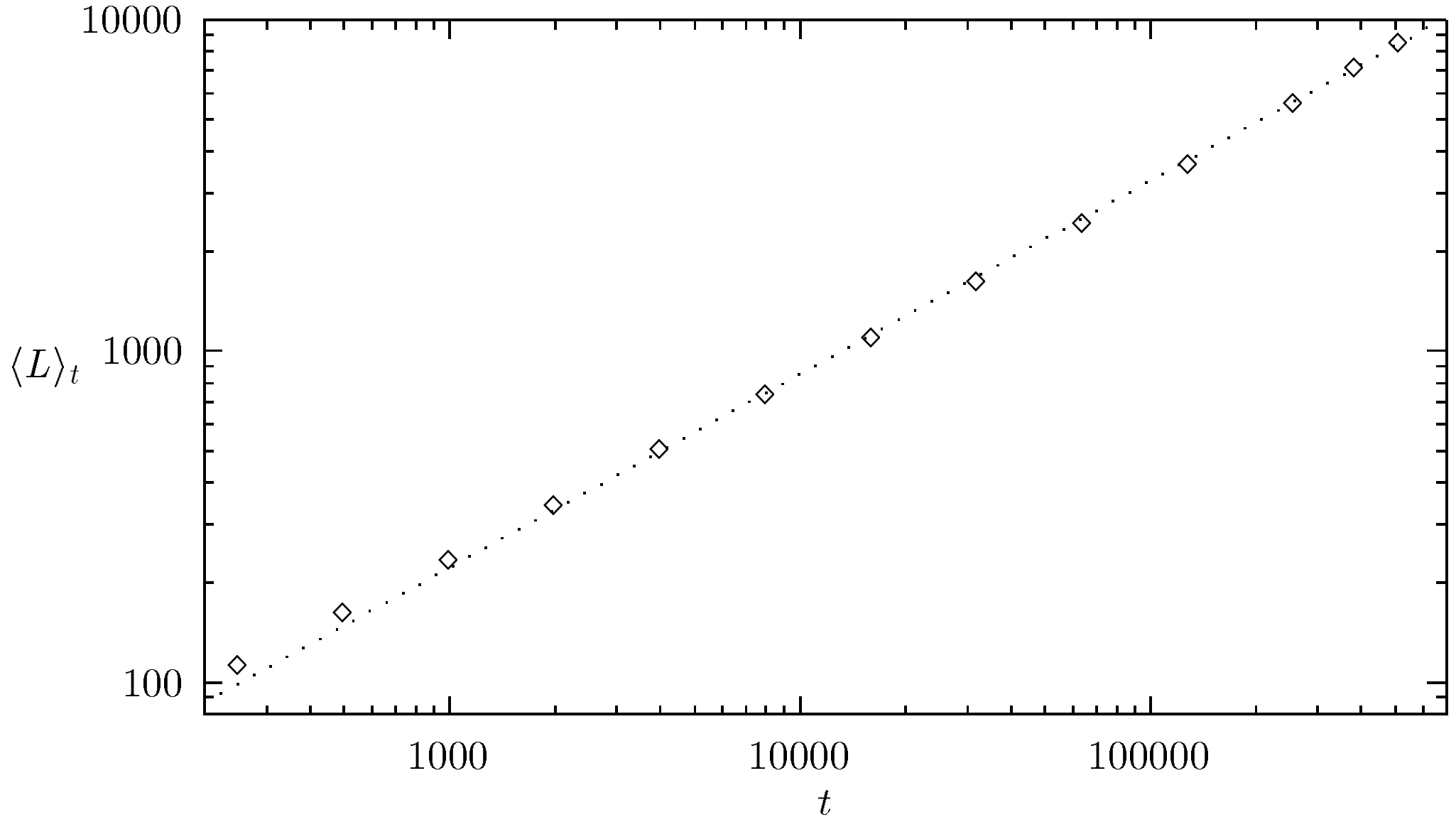

Figure 4: Log-log plot of the average domain size, $\langle L \rangle_t$. The dotted line has a slope of 7/12.

concentration evolves via

$$\frac{\partial \rho}{\partial t} \sim \frac{1}{\langle L \rangle \Delta t}\,, \tag{8}$$

where $\Delta t$ is the time it takes for a nearest-neighbor pair of opposite species to annihilate, and $\langle L \rangle$ is the typical cluster length.

If both members of the nearest-neighbor pair is assumed to diffuse,

$$\Delta t_{\rm diffuse} \sim \langle l_{AB} \rangle^2\,. \tag{9}$$

Note that this assumption of diffusive behavior is suspect: if we have an A-B pair, the A particle will move to the right with average velocity 1 and variance $\propto t$. The B particle will do the same only so long as it does not hit another B particle. If collisions of the B particle are relevant, and if the density immediately to the right of the B particle is $\rho_{\rm boundary}$, the B particle will move with average velocity $1 - \rho_{\rm boundary}$ and variance $\propto t^{2/3}$ [15]. Thus the A particle slowly catches up to the B particle in a time

$$\Delta t_{\rm drift} \sim \langle l_{AB} \rangle / [1 - (1 - \rho_{\rm boundary})] = \langle l_{AB} \rangle / \rho_{\rm boundary}\,, \tag{10}$$

provided that this is shorter than the diffusive time interval.

Plugging in our results for the behavior of $\langle l_{AB} \rangle$, and assuming that the boundary density is the same as the bulk, we have

$$\Delta t_{\rm diffuse} \sim t^{3/4} \qquad \Delta t_{\rm drift} \sim t^{17/24}\,, \tag{11}$$

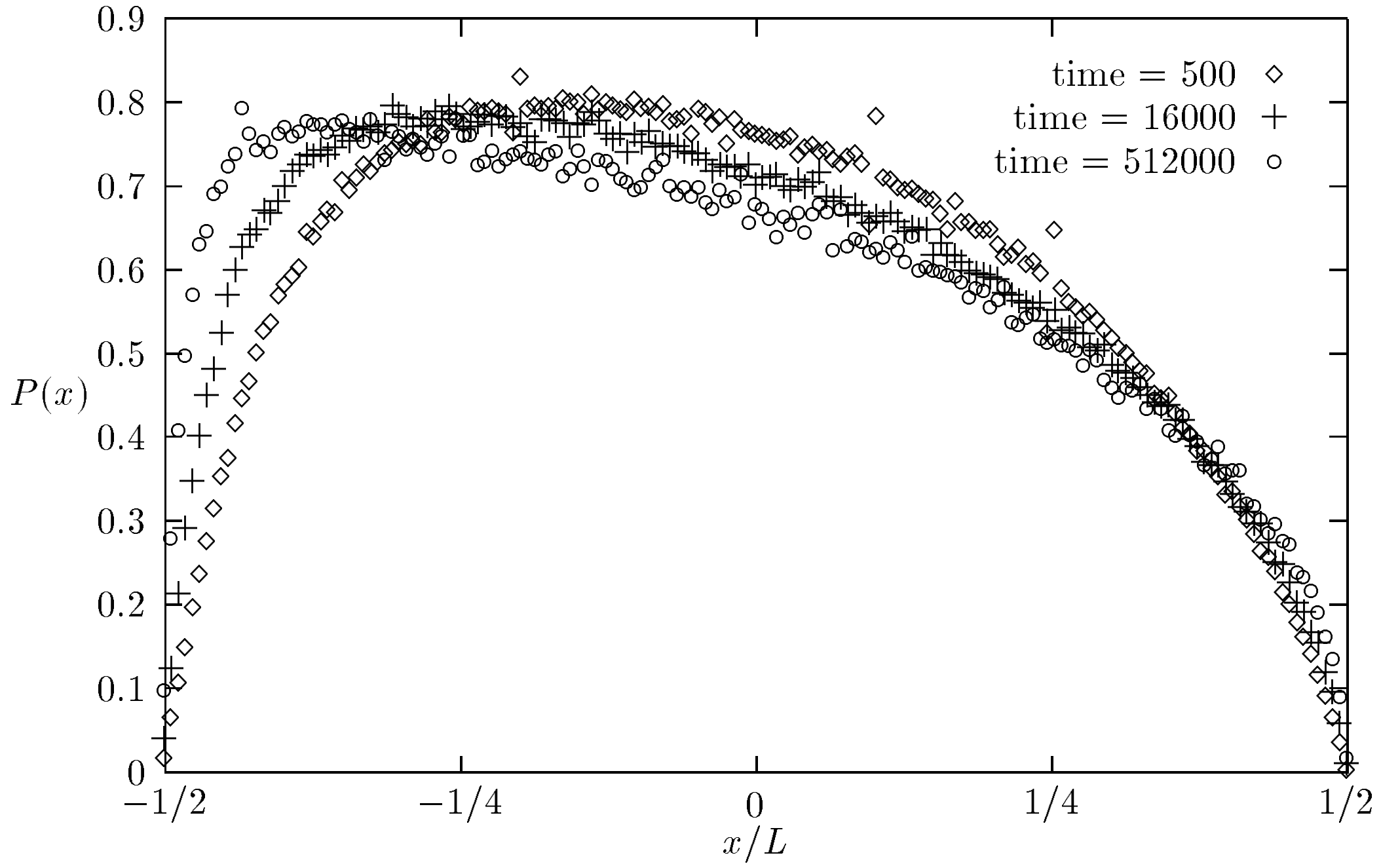

Figure 5: Scaled domain profile at various times. $L$ is the length of the cluster and $x$ is measured from the center of each cluster.

and thus, since $\dot{\rho} \sim t^{-4/3}$,

$$\langle L \rangle_t \sim t^{7/12} \text{ or } t^{5/8} \tag{12}$$

depending upon whether drift or diffusion is relevant. Which is correct? The data in figure 4 seems consistent with the diffusive picture, $t^{7/12}$, which can only occur if $\rho_{\text{boundary}} \ll \rho_{\text{bulk}}$.

How does $\rho_{\text{boundary}}$ behave? Is $\rho_{\text{boundary}} \ll \rho_{\text{bulk}}$, so that the observations and calculations are consistent? One way to approach this is to examine the density profile within a domain: in figure 5 we plot the probability that a particle in a cluster of length $L$ is a distance $x/L$ from the center of the cluster. Unlike the case with no drift [6], this probability does not seem to have a scaling form. At best one could say that near the center of the distribution one is approaching a scaling form. In particular the behavior near the boundaries seems to be getting sharper and sharper. This suggests that perhaps $x/L$ is not the correct scaling variable for cluster boundaries, *i.e.* the boundary behavior might scale differently from the bulk.

A first guess might be that at the boundary between clusters the A-A distance might in fact scale similarly to the A-B distance. To check this we re-plot the data in figure 5 with a new horizontal axis of $x/(Lt^{-5/24})$, with $x$ measured from the *left* edge of the cluster; this scaling is chosen because we want to rescale relative to

$$\langle l_{AB} \rangle_t \sim t^{3/8} = t^{7/12} t^{-5/24} \sim \langle L \rangle_t t^{-5/24}; \tag{13}$$

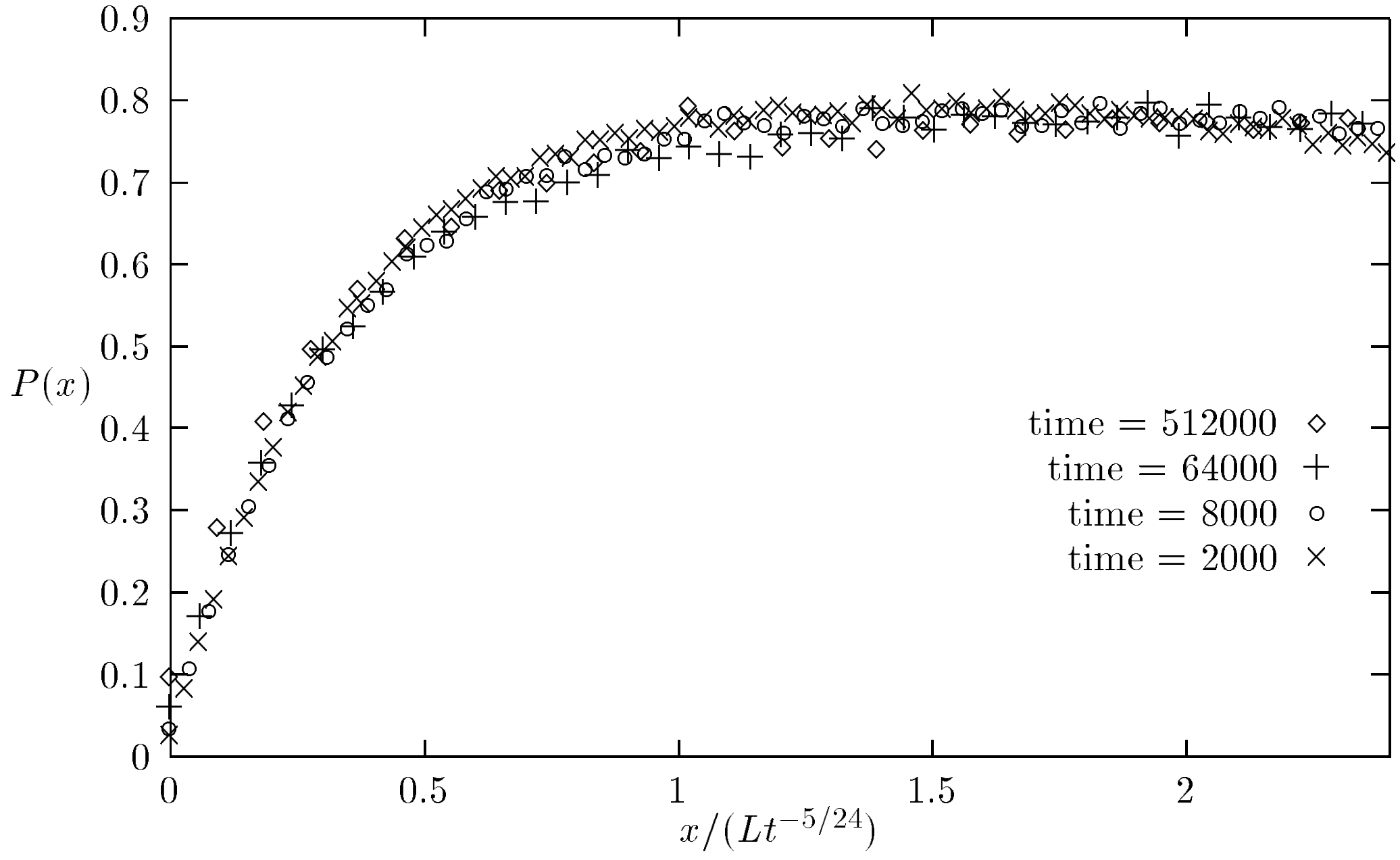

Figure 6: Scaled domain profile at various times. The position $x$ is measured from the left edge of each cluster.

when we do this in figure 6 we see that we obtain excellent scaling behavior over a very large range of times. This implies that $t^{3/8}$ is the correct length scale on the left-hand edge of the cluster, so that $\rho_{\mathrm{boundary}} \sim t^{-3/8}$, which when inserted into eq. (10) gives

$$\Delta t_{\mathrm{diffuse}} \sim \Delta t_{\mathrm{drift}} \sim t^{17/24}, \tag{14}$$

and thus the result that

$$\langle L \rangle_t \sim t^{7/12} \tag{15}$$

is consistent with both derivations.

If one instead looks at the profile near the right-hand edge of the cluster, neither $\langle L \rangle_t$ nor $\langle l_{AB} \rangle_t$ appears to be the appropriate length scale. Of course because of the asymmetry of the system, one should not necessarily expect that the two boundaries should behave similarly. Nevertheless, one might still expect there to be scaling; perhaps with a different relevant length: is there any type of scaling that is correct here?

The answer appears to be yes: although the scaling is certainly not as good as on the left-hand boundary, rescaling by $Lt^{-.12}$ seems to be appropriate), as seen in figure 7, where we plot the profile against $Lt^{-.12}$. (A naiive error analysis would place the rescaling length between $Lt^{-.1}$ and $Lt^{-.14}$.) While not entirely convincing, particularly given that we have no theoretical explanation for this result, it is not entirely surprising that there is some means of superposing the data from different times. So apparently there is a fourth relevant distance in this model, $t^{7/12}t^{-.12} = t^{.46}$.

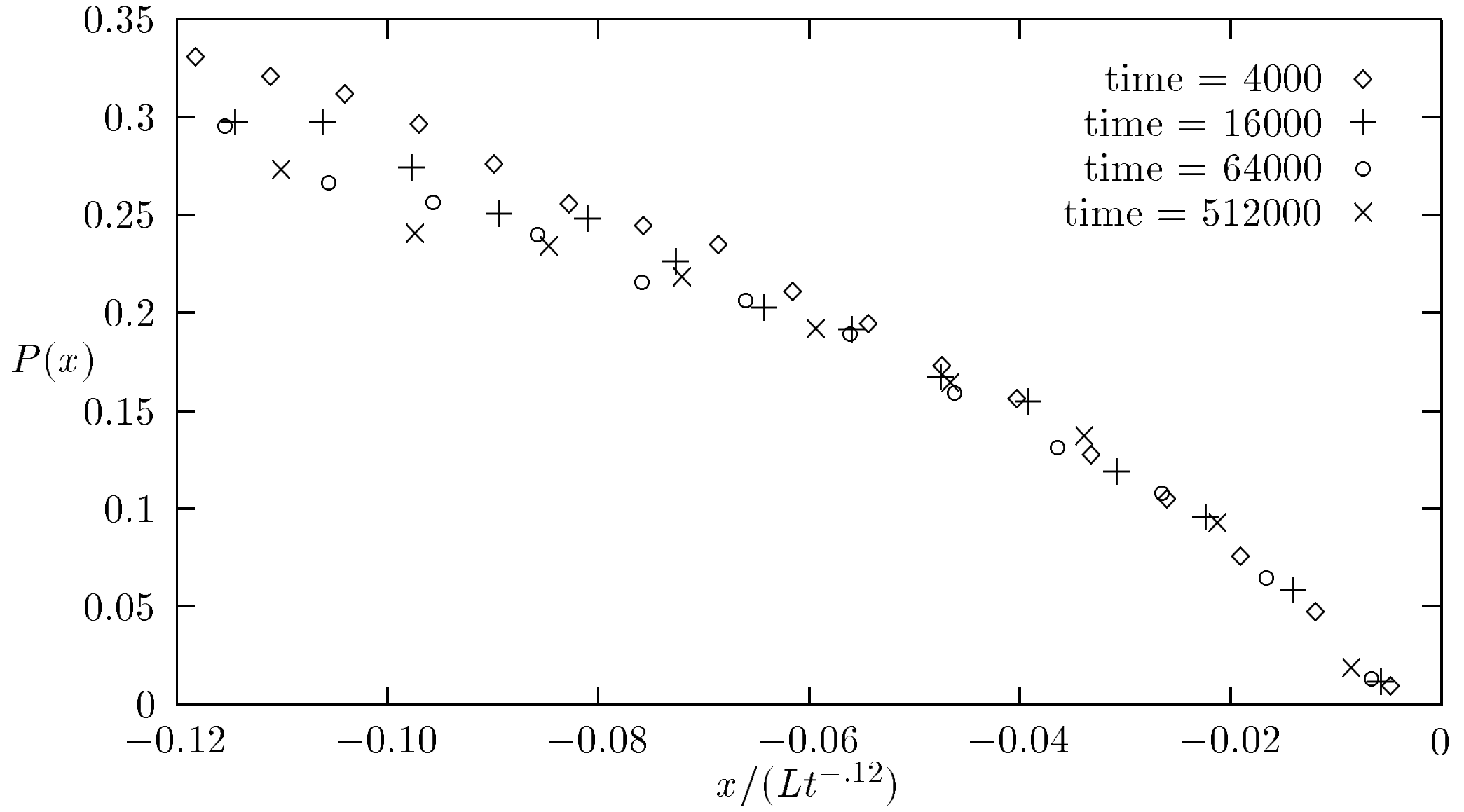

Figure 7: Scaled domain profile at various times. The position $x$ is measured from the right edge of each cluster.

# 4 Conclusion

In the ordinary, drift-free model for annihilating particles, there are three distinct (related) length scales. When a drift is added, we find four length scales.

The first and most important length scale is $\langle l_{AA}\rangle_t \sim t^{1/3}$. Since the clusters turn out to be large compared to the cluster separation, it is the nearest-neighbor same species particle distance that determines the density: $\rho \sim 1/\langle l_{AA}\rangle_t \sim t^{-1/3}$.

For the inter-cluster distance we find that $\langle l_{AB}\rangle_t \sim t^{3/8}$. This is the only length scale unchanged by the presence of the drift field.

The cluster size scales as $\langle L\rangle_t \sim t^{7/12}$. If this seems unlikely, recall that this is growth resulting from a combination of the expansion of individual clusters and the merging of pairs of clusters, so there is no reason for it to be particularly slow.

Lastly, we have the length scale $t^{.46}$ seen in the right-hand-side of each cluster. Because of symmetry, this length scale does not appear at all in the drift free case; its existence, as well as the different values for the exponents of the other length scales, indicates very clearly that the addition of a drift changes the universality class for the problem of annihilating particles.

# References

[1] D. Toussaint and F. Wilczek: "Particle-antiparticle annihilation in diffusive motion," *J. Chem. Phys.* **78**, 2642–2647 (1983).

[2] K. Kang and S. Redner: "Scaling Approach for the Kinetics of Recombination Processes," *Phys. Rev. Lett.* **52**, 955–958 (1984).

[3] K. Kang and S. Redner: "Fluctuation-dominated kinetics in diffusion–controlled reactions," *Phys. Rev. A* **32**, 435–447 (1985).

[4] M. Bramson and J. L. Lebowitz: "Asymptotic Behaviour of Densities for Two–Particle Annihilating Random Walks," *J. Stat. Phys.* **62**, 297–372 (1991).

[5] F. Leyvraz and S. Redner: "Spatial Organization in the Two-species Annihilation Reaction $A + B \to 0$," *Phys. Rev. Lett.* **66**, 2168–2171 (1991).

[6] S. Redner and F. Leyvraz: "Spatial Organization in Two-species Annihilation," *Phys. Rev. A* **46**, 3132 (1992).

[7] V. Belitsky: "Asymptotic Upper Bound of Density for Two-Particle Annihilating Exclusion," *J. Stat. Phys.* **73**, 671–694 (1993).

[8] S. A. Janowsky: "Asymptotic behavior of A + B → inert for particles with a drift," *Phys. Rev. E*, to appear.

[9] Y. Elskens and H. L. Frisch: "Annihilation kinetics in the one-dimensional ideal gas," *Phys. Rev. A* **31**, 3812–3816 (1985).

[10] V. Belitsky and P. Ferrari: "Ballistic Annihilation and Deterministic Surface Growth,"submitted to *J. Stat. Phys.*, São Paulo preprint (1994).

[11] H. van Beijeren, R. Kutner and H. Spohn: "Excess Noise for Driven Diffusive Systems," *Phys. Rev. Lett.* **54**, 2026–2029 (1985).

[12] L.-H. Gwa and H. Spohn: "Bethe Solution for the Dynamical Scaling Exponent of the Noisy Burgers Equation," *Phys. Rev. A* **46**, 844 (1992).

[13] H. Spohn: *Large-Scale Dynamics of Interacting Particles* Texts and Monographs in Physics, Springer-Verlag, 1991.

[14] V. Privman: "Discrete to Continuous-Time Crossover Due to Anisotropy in Diffusion-Limited Two-Particle Annihilation Reactions," *J. Stat. Phys.* **72**, 845–854 (1993).

[15] F. J. Alexander, S. A. Janowsky, J. L. Lebowitz and H. van Beijeren: "Shock Fluctuations in One-dimensional Lattice Fluids," *Phys. Rev. E* **47**, 403–410 (1993).